\documentclass[aps,prd,twocolumn,showpacs,amsmath,amssymb]{revtex4-1}
\usepackage{amsmath}
\usepackage{graphicx}
\usepackage{subfigure}
\usepackage{epstopdf}
\usepackage{color}
\usepackage{multirow}
\usepackage{setspace}
\usepackage{overpic}
\usepackage{amssymb}
\usepackage[bookmarksnumbered, pdfstartview=FitH,colorlinks,urlcolor=blue, citecolor=blue,linkcolor=blue] {hyperref}
\usepackage{lineno}
\usepackage{bm}
\usepackage{rotating}
\usepackage{longtable}
\usepackage{verbatim}
\usepackage{appendix}
\usepackage[utf8]{inputenc}

\let\oldequation\equation
\let\oldendequation\endequation

\renewenvironment{equation}
  {\linenomathNonumbers\oldequation}
  {\oldendequation\endlinenomath}

\begin{document}
%\linenumbers

\title{\bf \boldmath
Improved measurement of the branching fractions of the inclusive decays $D^+ \to K_S^0X $ and $D^0 \to K_S^0X $}

\author{
M.~Ablikim$^{1}$, M.~N.~Achasov$^{12,b}$, P.~Adlarson$^{72}$, M.~Albrecht$^{4}$, R.~Aliberti$^{33}$, A.~Amoroso$^{71A,71C}$, M.~R.~An$^{37}$, Q.~An$^{68,55}$, Y.~Bai$^{54}$, O.~Bakina$^{34}$
% ...
\\
\vspace{0.2cm}
(BESIII Collaboration)\\
\vspace{0.2cm} {\it
$^{1}$ Institute of High Energy Physics, Beijing 100049, People's Republic of China\\
% ...
$^{4}$ Bochum  Ruhr-University, D-44780 Bochum, Germany\\
% ...
$^{12}$ G.I. Budker Institute of Nuclear Physics SB RAS (BINP), Novosibirsk 630090, Russia\\
% ...
$^{33}$ Johannes Gutenberg University of Mainz, Johann-Joachim-Becher-Weg 45, D-55099 Mainz, Germany\\
$^{34}$ Joint Institute for Nuclear Research, 141980 Dubna, Moscow region, Russia\\
% ...
$^{37}$ Liaoning Normal University, Dalian 116029, People's Republic of China\\
% ...
$^{54}$ Southeast University, Nanjing 211100, People's Republic of China\\
$^{55}$ State Key Laboratory of Particle Detection and Electronics, Beijing 100049, Hefei 230026, People's Republic of China\\
% ...
$^{68}$ University of Science and Technology of China, Hefei 230026, People's Republic of China\\
% ...
$^{71}$ University of Turin and INFN, (A)University of Turin, I-10125, Turin, Italy; (B)University of Eastern Piedmont, I-15121, Alessandria, Italy; (C)INFN, I-10125, Turin, Italy\\
$^{72}$ Uppsala University, Box 516, SE-75120 Uppsala, Sweden\\
% ...
$^{b}$ Also at the Novosibirsk State University, Novosibirsk, 630090, Russia\\
% ...
}
}
%%% Local Variables:
%%% mode: latex
%%% TeX-master: "draft_BAM186"
%%% End:

\begin{abstract}
    By analyzing 2.93 fb$^{-1}$ of $e^+e^-$ collision data taken at the center-of-mass energy of 3.773 GeV with the BESIII detector,
    the branching fractions of the inclusive decays $D^+\to K^0_S X$
    and $D^0\to K^0_S X$
    are measured
    to be $(32.78\pm 0.13\pm 0.27)\%$ and $(20.54\pm 0.12\pm 0.18)\%$, respectively, where the first uncertainties are statistical and the second are systematic.
    These results are consistent with the world averages of previous measurements, but with improved precision.

\end{abstract}


\maketitle

\oddsidemargin  -0.2cm
\evensidemargin -0.2cm

\section{Introduction}

Weak decays of charmed mesons offer an ideal laboratory for improving our understanding of both strong
and weak interactions. The $D^+$ and $D^0$ mesons decay dominantly into final states involving $K^-$ and $\bar K^0$ via Cabibbo-favored processes.
The branching fractions of the inclusive decays $D^{+(0)}\to K^0/\bar K^0 X$, where $X$ denotes any possible particle combinations, were first measured by Mark-III~\cite{Mark3-KSX} in 1991
and later by BES~\cite{BESII-KSX} in 2006.
Assuming $\mathcal B(K^0/\bar K^0\to K^0_S)=0.5$ and considering the world average branching fractions of $D^{+(0)}\to K^0/\bar K^0 X$, the branching fractions of $D^{+(0)}\to K^0_S X$ are expected to be $(30.5\pm2.5)\%$ and $(23.5\pm2.0)\%$, respectively~\cite{pdg2020}.
In recent years,
the precision of the branching fractions of known decay modes has been significantly improved, and
many new channels have been observed.
Following this progress, the sum of the branching fractions of the exclusive $D^+$ and $D^0$ decays involving $K^0_S$ are found to be $(31.68\pm0.32)$\% and $(18.16\pm0.72)$\%,
as summarized in Appendix~\ref{D0Dpmodes}.
In contrast, however, the branching fractions of the inclusive decays $D^{+}\to K^0_S X$ and $D^{0}\to K^0_S X$ suffer from large uncertainties due to the limited samples sizes used to perform the measurements.
Improved determinations of these inclusive decay branching fractions will help to quantify and guide the search for the missing decay modes involving a $K^0_S$ meson.

In this paper, we report the improved measurements of the branching fractions of $D^{+}\to K^0_S X$ and $D^{0}\to K^0_S X$, obtained from the analysis of  $2.93\ \text{fb}^{-1}$ of $e^+e^-$ collision data~\cite{lum_bes3} taken at the center-of-mass energy $\sqrt{s}$ of 3.773 GeV with the BESIII detector. The charge-conjugated processes are implied throughout the discussion.

\section{BESIII detector and Monte Carlo simulation}

The BESIII detector~\cite{BESIII} records symmetric $e^+e^-$ collisions
provided by the BEPCII storage ring~\cite{Yu:IPAC2016-TUYA01}, which operates in the center-of-mass energy range from 2.0 to 4.95~GeV, with a peak luminosity of $1\times10^{33}$~cm$^{-2}$s$^{-1}$ achieved at $\sqrt{s}=3.773$ GeV.
BESIII has collected large data samples in this energy region~\cite{Ablikim:2019hff}. The cylindrical core of the BESIII detector covers 93\% of the full solid angle and consists of a helium-based
 multilayer drift chamber~(MDC), a plastic scintillator time-of-flight
system~(TOF), and a CsI(Tl) electromagnetic calorimeter~(EMC),
which are all enclosed in a superconducting solenoidal magnet
providing a 1.0~T magnetic field.~\cite{detvis} The solenoid is supported by an
octagonal flux-return yoke with resistive plate counter muon
identification modules interleaved with steel.
The charged-particle momentum resolution at $1~{\rm GeV}/c$ is
$0.5\%$, and the specific energy loss~(${\rm d}E/{\rm d}x$) resolution is $6\%$ for electrons
from Bhabha scattering. The EMC measures photon energies with a
resolution of $2.5\%$ ($5\%$) at $1$~GeV in the barrel (end-cap)
region. The time resolution in the TOF barrel region is 68~ps, while
that in the end-cap region is 110~ps.

Simulated samples produced with the {\sc geant4}-based~\cite{geant4} Monte Carlo (MC) package, which
includes the geometric description of the BESIII detector and the
detector response, are used to determine the detection efficiency
and to estimate the backgrounds. The simulation includes the beam-energy spread and initial-state radiation in the $e^+e^-$
annihilations modeled with the generator {\sc kkmc}~\cite{kkmc}.
The inclusive MC samples consist of the production of $D\bar{D}$
pairs with consideration of quantum coherence for all neutral $D$
modes, the non-$D\bar{D}$ decays of the $\psi(3770)$, the initial-state radiation
production of the $J/\psi$ and $\psi(3686)$ states, and the
continuum processes.
The known decay modes are modeled with {\sc
evtgen}~\cite{evtgen} using the branching fractions taken from Particle Data Group (PDG)~\cite{pdg2020}, and the remaining unknown decays of the charmonium states are
modeled by {\sc
lundcharm}~\cite{lundcharm}. Final-state radiation from charged final-state particles is incorporated using the {\sc
photos} package~\cite{photos}.

The three-body decays $D^+\to K_{S}^{0}\pi^{+}\pi^{0}$ and $D^+\to K_{S}^{0}\pi^{+}\eta$ are modeled with
 a modified data-driven generator BODY3~\cite{evtgen}, which was developed to simulate different intermediate states in data for a given three-body final state.  The Dalitz plot from data, corrected for backgrounds and efficiencies, is taken as input for the BODY3 generator.
The decay $D^+\to K_{S}^{0}\pi^+\pi^+\pi^-$ is modeled with the partial-wave analysis model from Ref.~\cite{bes3-kspipipi}.
Other decays reported in recent BESIII measurements~\cite{bes3-D-KKpipi,ksompi,lanxing} are modeled by
constructing a cocktail of possible intermediate resonances.

\section{Analysis Method}

At $\sqrt s=3.773$~GeV, $D^0\bar D^0$ or $D^+D^-$ pairs are produced without accompanying hadron(s),
thereby offering a clean environment in which to investigate hadronic $D$ decays with the double-tag (DT) method~\cite{Li:2021iwf}.
To measure the branching fractions of the hadronic decays of $D^+$($D^0$) mesons,
as a first step the single-tag $D^-$($\bar D^0$) mesons are selected by using the hadronic decay modes $D^-\to K^+\pi^-\pi^-$ and $\bar D^0\to K^+\pi^-$.
The number of the expected single-tag $D^-$ ($\bar D^0$) mesons is given by
\begin{equation}
\label{equ:singlytag}
  N_{\rm tag} = 2N_{{D \bar D}}{\mathcal B}_{\rm tag}\epsilon_{\rm tag},
\end{equation}
where $N_{{D \bar D}}$ is the total number of
$D^{+}D^{-}$ ($D^0\bar D^0$) pairs,
${\mathcal B}_{\rm tag}$ is the branching fraction of $D^-$ ($\bar D^0$) decay into the tag mode,
and $\epsilon_{\rm tag}$ is the efficiency of reconstructing the single-tag $D^-$ ($\bar D^0$) mesons.

The double-tag sample is comprised of events in which the signal decay $D^{+(0)}\to K_S^0X$ can be selected in the presence of the single-tag $D^-$ ($\bar D^0$) mesons. The number of the expected double-tag events is given by
\begin{equation}\label{equ:doubletag}
  N_{\rm DT} =
2N_{D\bar D} {\mathcal B}_{\rm tag} {\mathcal B}_{{\rm sig}}
\epsilon_{{\rm DT}},
\end{equation}
where ${\mathcal B}_{{\rm sig}}$
is the branching fraction of $D^{+(0)}\to K_S^0X$,
and $\epsilon_{{\rm DT}}$
is the efficiency of detecting the double-tag events. The branching fraction of the signal decay is then
\begin{equation}\label{equ:doubletagB}
  {\mathcal B}_{{\rm sig}} =
\frac{N_{{\rm DT}}}
{N_{\rm tag}\epsilon_{{\rm sig}}},
\end{equation}
where $\epsilon_{{\rm sig}} =
\frac{\epsilon_{{\rm DT}}}{\epsilon_{\rm tag}}$ is the effective efficiency of reconstructing $D^{+(0)}\to K_S^0X$.

Quantum correlations between the $D^0 \bar{D^0}$ pair modify the double-tag rates.
The measured branching fraction of $D^0\to K^0_SX$ tagged by $\bar D^0\to K^+\pi^-$ must be corrected to
\begin{equation}\label{equ:QCcorrect}
{\mathcal B}^{\rm cor}_{\rm sig} = \frac{1}{1-C_f(2f_{\rm CP+}-1)}{\mathcal B}_{\rm sig},
\end{equation}
where  $f_{CP+}$ is the fractional $CP$-even content of the decay $D^0\to K^0_SX$, and $C_{f}$ is the strong-phase factor
\begin{equation}
C_{f} = \frac{2r \cos\delta}{1 + r^2}.
\end{equation}
Here $r=0.0587\pm0.0002$~\cite{A1} is the ratio between doubly-Cabibbo-suppressed and Cabibbo-favored amplitudes for $D^0/\bar D^0 \to K^\pm\pi^\mp$, and
$\delta=(191.7_{-3.8}^{+3.6})^\circ$~\cite{A1} is the strong-phase difference between two amplitudes~\cite{refcp3, refcp4}.
These inputs yield $C_f=(-11.5\pm0.2)\%$ for $D^0/\bar D^0 \to K^\pm\pi^\mp$.

The $CP$-even fraction $f_{CP+}$ of $D^0\to K^0_SX$ decays is not known, but can be measured from the $D^0\bar{D^0}$ data set. Following Refs.~\cite{refcp1,refcp2},
the $CP$-even fraction  is calculated by
\begin{equation}
f_{\rm CP\pm} = \frac{N^{\pm}}{N^++N^-},
\nonumber
\end{equation}
\begin{equation}
N^{\pm} = \frac{M^{\pm}_{\rm measured}}{S^{\pm}},
\nonumber
\end{equation}
and
\begin{equation}
S_{\pm} = \frac{S^{\pm}_{\rm measured}}{1 - \eta_{\pm}y_D},
\nonumber
\end{equation}
where $N^{\pm}$ is the ratio between double-tag and single-tag yields with $CP$-odd(even) tags,
$M_{\rm measured}^{\pm}$ is the number of double-tag candidates for a signal channel versus CP-odd(even) tags,
$S^{\pm}_{\rm measured}$ is the number of single-tag candidates for $CP$-even(odd) decay modes,
$S^{\pm}$ is the number of the corrected single-tag candidates for $CP$-even(odd) decay modes.
Finally, $\eta_\pm=\pm 1$ for $CP$-even(odd) mode and $y_D=(6.47\pm0.24)\times10^{-3}$ is the mixing parameter taken from the latest average from HFLAV~\cite{A1}.

\section{Yields of single-tag $D^-\,(\bar D^0)$ mesons}

Charged kaons  and charged pions  are selected with the same selection criteria as those used in our previous works~\cite{bes3-pimuv,bes3-etaetapi,bes3-etaX}.
All charged tracks, apart from those from $K^0_{S}$ decays, are
required to have a polar angle $\theta$ with respect to the MDC
symmetry axis
within the MDC acceptance $|\rm{cos\theta}|<0.93$, a distance of closest approach with respect to the interaction point along the beam direction $|V_{z}|<$ 10 cm, and a distance of closest approach with respect to the interaction point in the plane transverse to the beam direction $|V_{xy}|<$ 1 cm. Particle identification (PID) for charged kaons and pions is performed by
exploiting combined d$E/$d$x$ and TOF information.
The confidence levels for pion and kaon hypotheses ($CL_{\pi}$ and $CL_{K}$) are calculated. Charged tracks satisfying $CL_{K}>CL_{\pi}$ and $CL_{\pi}>CL_{K}$ are assigned as charged kaons and pions, respectively.

When selecting $\bar D^0\to K^+\pi^-$ candidates, contamination from cosmic rays and Bhabha events is suppressed with the following requirements~\cite{deltakpi}. First, the two charged tracks must have a TOF time difference less than 5~ns and they must be inconsistent with being a muon-antimuon or an electron–positron pair. Second, in the event there must be at least one EMC shower with an energy greater than 50 MeV or at least one additional charged track detected in the MDC.

The tagged $D^-\,(\bar D^0)$ mesons are identified using the energy difference $\Delta E \equiv E_{\bar D} - E_{\rm beam}$ and the beam-constrained mass $M_{\rm BC} \equiv \sqrt{E^{2}_{\rm beam}-|\vec{p}_{\bar D}|^{2}}$.
Here, $E_{\rm beam}$ is the beam energy, and
$\vec{p}_{\bar D}$ and $E_{\bar D}$ are the momentum and energy of
the $D^-\,(\bar D^0)$ candidate in the rest frame of $e^+e^-$ system, respectively.
For each tag mode, if there are multiple candidates in an event,
only the one with the smallest $|\Delta E|$ is kept.
The tagged $\bar D$ candidates are required to be within
$\Delta E\in(-25,25)$\,MeV.

\begin{figure}[htp]
  \centering
\includegraphics[width=1.0\linewidth]{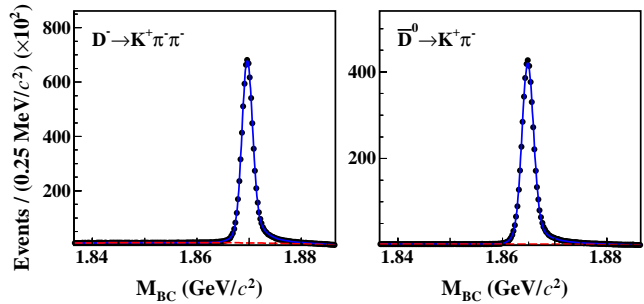}
  \caption{
Best fits to the $M_{\rm BC}$ distributions of the single-tag candidates for $D^-\to K^+\pi^-\pi^-$ and $\bar D^0\to K^+\pi^-$. The points with error bars are data. The blue solid curves are the fit results. The red dashed curves are the fitted combinatorial background shapes.}
\label{fig:datafit_tag_MassBC}
\end{figure}

To determine the yields of single-tag $D^-\,(\bar D^0)$ mesons for individual tag modes, binned maximum-likelihood fits are performed on the corresponding $M_{\rm BC}$
distributions of the accepted single-tag candidates~\cite{bes3-pimuv,bes3-etaetapi,bes3-etaX}.
In the fits, the $D^-\,(\bar D^0)$ signal is modeled by an MC-simulated shape convolved with
a double-Gaussian function describing the resolution difference between data and MC simulation.
The combinatorial background is described by an ARGUS function~\cite{ARGUS}.
The best fit results for the $M_{\rm BC}$ distributions are shown in Fig.~\ref{fig:datafit_tag_MassBC}.
The yields of the single-tag $D^-$ and $\bar D^0$ mesons are $798935\pm1011$ and $529227\pm761$ respectively, where the uncertainties are statistical only.
The efficiencies of reconstructing the single-tag $D^-\,(\bar D^0)$ mesons are estimated
to be $(51.90\pm0.08)\%$  and $(65.60\pm0.09)\%$ respectively,
by analyzing the inclusive MC sample.

\section{Yields of double-tag $D^{+(0)}$ mesons}

\begin{figure}[htp]
\centering
\includegraphics[width=1.0\linewidth]{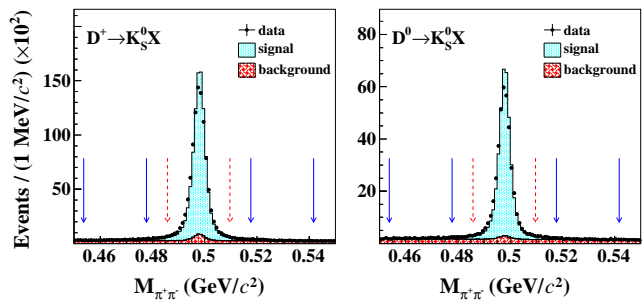}
  \caption{
  Distributions of $M_{\pi^+\pi^-}$ of the candidates for $D^+\to K^0_SX$ and $D^0\to K^0_SX$.
  The red dashed and blue solid arrow pairs show the $K^0_S$ signal and sideband regions, respectively.
  The points with error bars are data. The histograms are the inclusive MC sample.}
\label{fig:kosm}
\end{figure}

The $K^{0}_{S}$ mesons are reconstructed in the
$K^{0}_{S}\to\pi^{+}\pi^{-}$ decay mode.
The charged pions making up the $K^0_{S}$ candidates are required to satisfy $|V_{z}|<$ 20 cm and $|\!\cos\theta|<$ 0.93.
To improve the resolution of the  $K^{0}_{S}$ invariant mass,
a primary vertex and a $K^{0}_{S}$ decay vertex fit are performed on the $\pi^{+}\pi^{-}$ pairs.
To suppress the combinatorial background, we require
$L/\sigma_{L}>$ 2, where $\sigma_{L}$ is the uncertainty of the flight distance $L$.
The invariant mass of the chosen charged pions, $M_{\pi^+\pi^-}$, is required to be within $|M_{\pi^{+}\pi^{-}} - M_{K_{S}^{0}}|<$ 12 MeV/$c^{2}$,
which corresponds to about $\pm3\sigma$ around the known $K^0_S$ mass ($M_{K_{S}^{0}}$)~\cite{pdg2020}.
To estimate the combinatorial $\pi^+\pi^-$ background, the $K^{0}_{S}$ sideband region is defined by $20<|M_{\pi^{+}\pi^{-}} - M_{K_{S}^{0}}|<$ 44 MeV$/c^{2}$, which corresponds to about $\pm5\sigma$ to $\pm11\sigma$ away from the known $K^0_S$ mass.
The definitions of the $K^0_S$ signal and sideband regions are indicated in Fig~\ref{fig:kosm}. The background is mainly from the $D\bar D$ and $q\bar q$~($q = u,d,s$) continuum process. Especially, events with wrong tag and one real $K^0_S$ meson will form a peak in the $M_{\pi^+\pi^-}$ distribution but do not form a peak in the $M_{\rm BC}$ distribution. There are some discrepancy in resolution between data and MC simulation and this effect will be taken into account in the systematic uncertainty of $K_S^0$ reconstruction.

\begin{figure*}[htbp]
\centering
\includegraphics[width=1.0\textwidth]{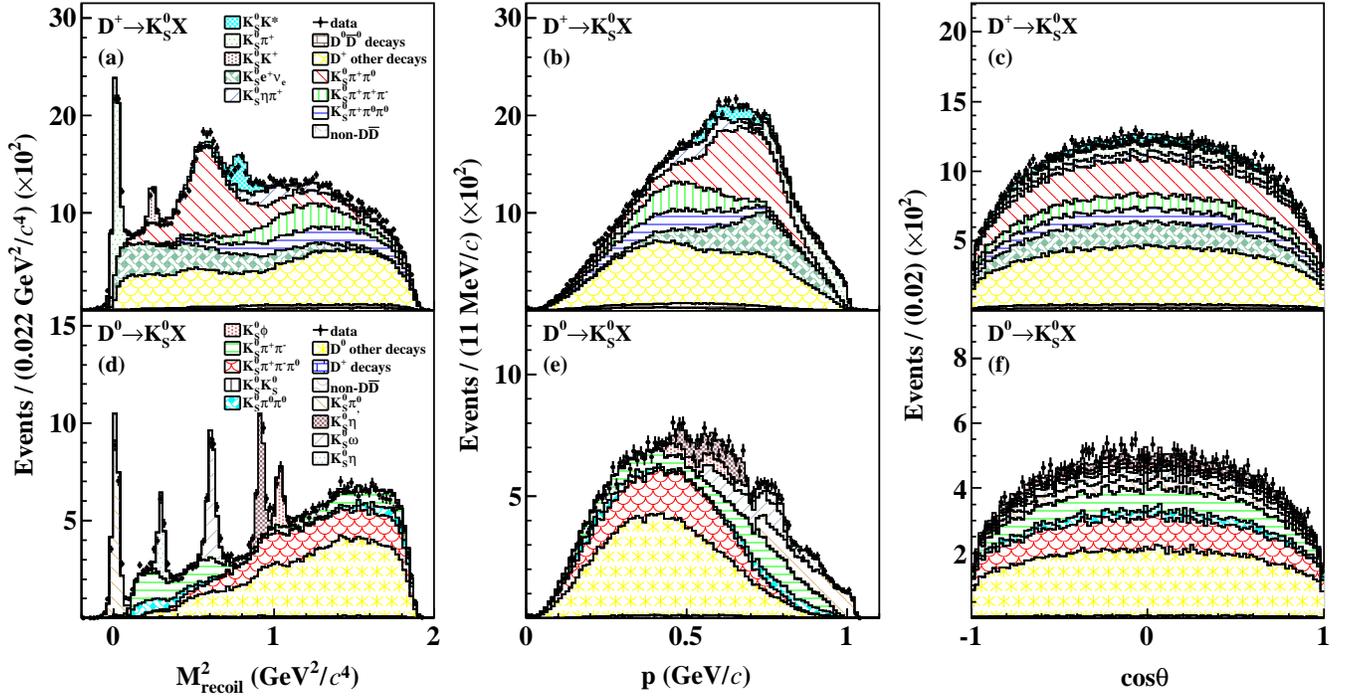}
\caption{Distributions of the (a,d) recoil mass, (b,e) momentum, and (c,f) $\cos\theta$ of the $K^0_S$ candidates from the accepted candidates for $D^{+}\to K_{S}^{0}X$ and $D^{0}\to K_{S}^{0}X$.
The points with error bars are data.
The histograms are the inclusive MC sample.}
\label{fig:DpD0}
\end{figure*}

The distributions of the recoil mass, momentum, and $\cos\theta$ of the $K^0_S$ candidates  from the selected $D^{+}\to K_{S}^{0}X$ and $D^{0}\to K_{S}^{0}X$ events
are shown in Fig.~\ref{fig:DpD0}. The experimental and the MC distributions are in good agreement.

\begin{figure}[htp]
  \centering
\includegraphics[width=1.0\linewidth]{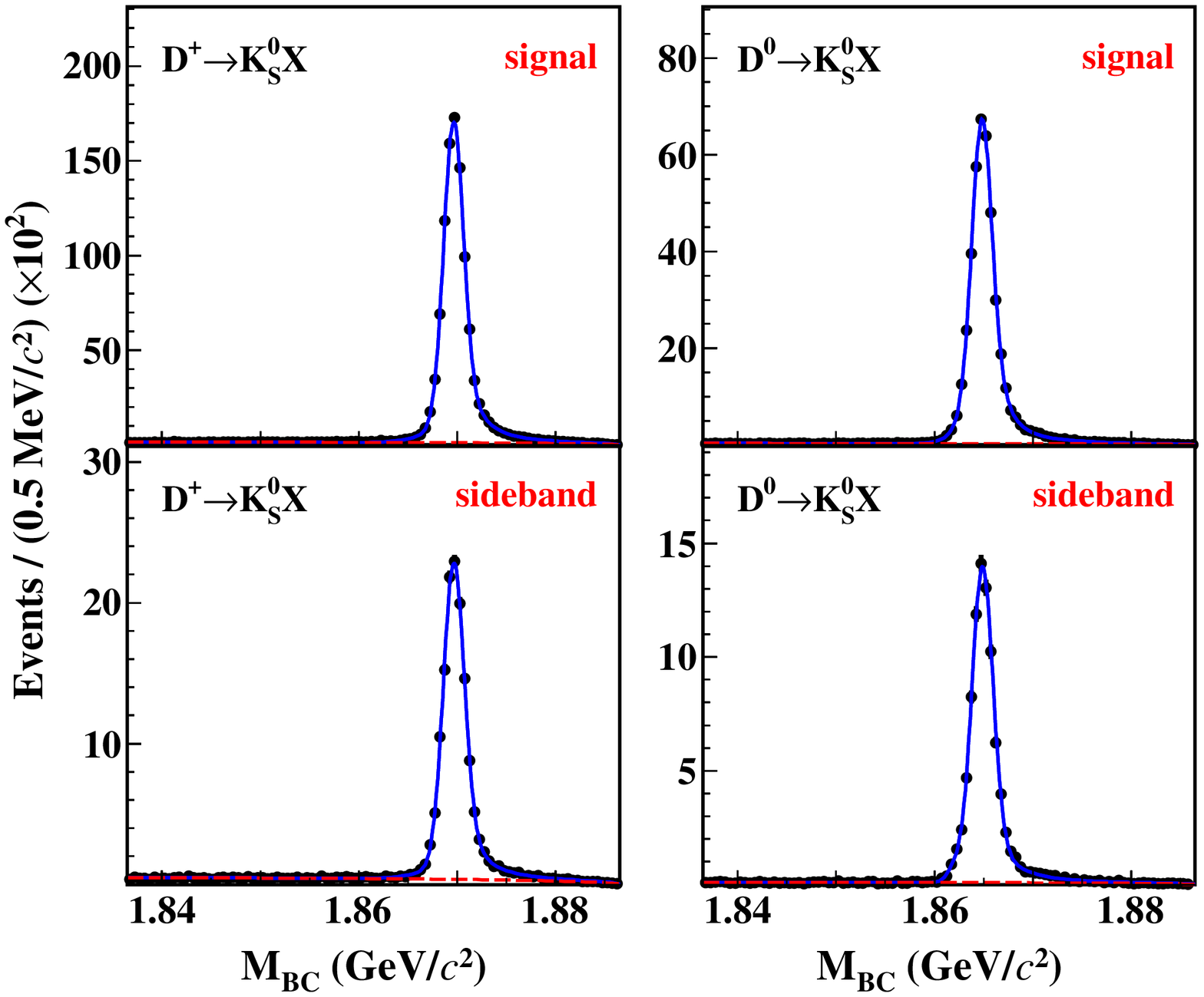}
  \caption{Best fits to the $M_{\rm BC}$ distributions of the double-tag events in data.
The upper and lower plots correspond to events with $M_{\pi^+\pi^-}$ in the $K^0_S$ signal and sideband regions, respectively.
The points with error bars are data, the blue solid curves are the fit results,
the red dashed curves are the fitted combinatorial background shapes.}
\label{fig:datafit_sig_MassBC}
\end{figure}

The decay yields of the signal are determined from the fits to the $M_{\rm BC}$ distributions of the accepted double-tag events in data.
However, combinatorial background events with $M_{\pi^+\pi^-}$ in the $K^0_S$ signal region may form peaking background in the $M_{\rm BC}$ distribution. So, the net signal yields are determined by
\begin{equation}
N_{\rm DT}^{\rm net}=N_{K^0_S\rm sig}-f_{\rm co} \cdot N_{K^0_S\rm sb}.
\label{eq:br}
\end{equation}
where $N_{K^0_S\rm sig}$ and $N_{K^0_S\rm sb}$ are the fitted double-tag yields in the $K^0_S$ signal and sideband regions, respectively;
the sideband scaling factor $f_{\rm co}$ is taken as 0.5 under the assumption that the background contribution in the $M_{\pi^+\pi^-}$ spectrum is flat.
Figure \ref{fig:datafit_sig_MassBC} shows the $M_{\rm BC}$ distributions of the accepted double-tag events in data with $M_{\pi^+\pi^-}$ in the $K^0_S$ signal and sideband regions.
 We perform similar fits to these spectra as for the fits to the single-tag candidates.
In these fits, the parameters of the smeared double-Gaussian resolution functions are fixed at the values obtained from the fits to the single-tag $D^-\,(\bar D^0)$ candidates. The fit results are also shown in Fig.~\ref{fig:datafit_sig_MassBC}. From these fits, we obtain the yields of $N_{K^0_S\rm sig}$ and $N_{K^0_S\rm sb}$, as summarized in Table~\ref{new BF}.

The efficiencies of reconstructing the inclusive decays $D^{+(0)}\to K_S^0X$ are determined by analyzing the inclusive MC sample~($\epsilon_{\rm sig}$).
Inserting the numbers of $N_{\rm DT}^{\rm net}$, $N_{\rm tag}$, and $\epsilon_{\rm sig}$ into Eq. (\ref{equ:doubletagB}),
we obtain the branching fractions of $D^{+}\to K_{S}^{0}X$  and $D^{0}\to K_{S}^{0}X$.

In order to account for the effect of quantum correlations on the branching fraction of $D^0 \to  K_S^0 X$, we apply the correction given in Eq.~\ref{equ:QCcorrect}.  This procedure requires knowledge of the $CP$-even fraction of these decays, which is measured in the data by using  the $CP$-even tag $\bar D^0\to K^+K^-$ and the CP-odd tag $\bar D^0\to K^0_S\pi^0$.  The yields of these tags and the resulting value for $f_{CP+}$ are given in Table~\ref{CP_pam}.
The correction factor  is determined to be $1.0204\pm0.0024$.

\begin{table}[htbp]
\centering
\caption{
Measurements of $S^\pm_{\rm measured}$, $M^\pm_{\rm measured}$, $f_{\rm CP+}$ and the correction factor
for $D^0\to K^0_SX$ due to quantum correlation effect. }
\label{CP_pam}
\centering
\begin{tabular}{ccc}
  \hline\hline
CP tag mode&$\bar D^0\to K^+K^-$&$\bar D^0\to K^0_S\pi^0$\\
  \hline
$S^\pm_{\rm measured}$ &$57779\pm287$ &$70512\pm311$\\
$M^\pm_{\rm measured}$ &$4760\pm81$  &$4068\pm78$\\
$f_{\rm CP+}$     &\multicolumn{2}{c}{$0.413\pm0.010$}\\
Correction factor &\multicolumn{2}{c}{$1.0204\pm0.0024$}\\
\hline\hline
\end{tabular}
\end{table}

The net efficiencies, including all factors discussed above, and the final values for the branching fractions are summarized in Table \ref{new BF}.

\begin{table*}[htbp]
\centering
\caption{
Numbers of single-tag $D^-\,(\bar D^0)$ mesons ($N_{\rm tag}$),
 fitted double-tag yields in the $K^0_S$ signal and sideband regions ($N_{K^0_S\rm sig}$ and $N_{K^0_S\rm sid}$),
 net numbers of signal events ($N_{\rm DT}^{\rm net}$),
 efficiencies of finding $D^{+(0)}\to K_{S}^{0}X$ in the presence of the single-tag $D^-\,(\bar D^0)$ meson ($\epsilon_{\rm sig}$), and
 measured branching fractions of $D^{+(0)}\to K_{S}^{0}X$ ($\mathcal{B}_{\rm sig}$).
The efficiency values include the branching fraction of $K^0_S\to \pi^+\pi^-$ as well as the correction factors described in the text.  The uncertainties are statistical only.
}
\begin{center}
\begin{tabular}{ccccccc}
\hline
\hline
 Signal mode   & $N_{\rm tag}$ &$N_{K^0_S\rm sig}$   &$N_{K^0_S\rm sb}$ & $N_{\rm DT}^{\rm net}$  &$\epsilon_{\rm sig}$  (\%) &$\mathcal{B}_{\rm sig}$  (\%)\\
\hline
$D^+ \to K_S^{0}X$  &$798935\pm1011$  &$102186\pm365$  &$13981\pm140$ &$95195\pm372$ &$36.35\pm0.08$ &$32.78\pm0.13$ \\
$D^0 \to K_S^{0}X$   &$529227\pm761$  &$42166\pm215$  &$8688\pm97$   &$37822\pm220$   &$34.81\pm0.10$ &$20.54\pm0.12$ \\

\hline
\hline
\end{tabular}
\end{center}
\label{new BF}
\end{table*}

\section{Systematic uncertainties}

With the double-tag method, the branching-fraction measurements have a
greatly reduced sensitivity to systematic bias arising from the selection criteria on the tag side.

The systematic uncertainties in the yields of the single-tag $D^-$ or $\bar D^0$ mesons have been estimated in Refs. \cite{bes3-pimuv} to be 0.50\%, by examining the relative change between data and MC simulation yields when varying the fit range, signal shape, and end point of the ARGUS function.

The data/MC difference of the $K^0_S$ reconstruction is determined to be
$(1.01\pm0.53)\%$ by using control samples of $J/\psi\to K^0_SK^\pm\pi^\mp$ and $\phi K^0_SK^\pm\pi^\mp$ decays~\cite{kso_reconstruction}. After correcting the MC efficiencies to match those in data, we assign 0.53\% as the systematic uncertainty per $K_S^0$ decay.

Two methods are used to estimate the systematic uncertainty associated with the $K_{S}^{0}$ multiplicity.
Firstly, we vary the components of the decays containing one, two, and three $K^0_S$ mesons by $\pm 1\sigma$ according to
their known branching fractions. The changes in the detection efficiencies for $D^+\to K^0_SX$ and $D^0\to K^0_SX$, 0.05\% and 0.07\% respectively, are assigned as the uncertainties.
Secondly, we measure the detection efficiencies for all known exclusive decays
and determine the change of the re-weighted detection efficiency by varying each of
decays within $\pm 1\sigma$ according to the known branching fractions.
By adding in quadrature all the changes in the detection efficiencies, we obtain
0.20\% for $D^+\to K^0_SX$ and 0.28\% for $D^0\to K^0_SX$, which
are assigned as the uncertainties.
We assign the larger effect coming from the two methods as the corresponding systematic uncertainty for each signal decay.

To estimate the systematic uncertainty due to the choice of the $K^0_S$ sideband region, we examine the effects on the branching fractions
by shifting the $K^0_S$ sideband region by $\pm2$ or $\pm 4$ MeV/$c^2$.
If the difference of the re-measured branching fraction is greater than the statistical uncertainty after considering the signal correlations, it is assigned as the systematic uncertainty; otherwise, it is  neglected. Following this procedure, we assign 0.15\% for $D^+\to K_S^0X$ and 0.10\% for $D^0\to K_S^0X$ as the corresponding systematic uncertainties.

The uncertainty on the correction factor for quantum correlations is propagated to the branching fraction of $D^0 \to  K_S^0 X$ decays.
The uncertainties due to the limited signal double-tag MC samples for $D^{+}\to K_{S}^{0}X$ and $D^{0}\to K_{S}^{0}X$ are estimated to be 0.25\% and 0.28\%, respectively.
The uncertainty of the branching fraction of $K^0_S\to \pi^+\pi^-$ is  0.07\%, taken from the PDG~\cite{pdg2020}.

The uncertainty arising from the scale factor $f_{\rm co}$ used in Eq.~(\ref{eq:br}) is assessed with the alternative scale factors derived from the inclusive MC sample,
which are 0.549 and 0.534 for $D^{+}\to K_{S}^{0}X$ and $D^{0}\to K_{S}^{0}X$, respectively. The changes of the re-measured branching fractions,
0.02\% and 0.05\%, are assigned as the corresponding systematic uncertainties.

The systematic uncertainties are summarized in Table~\ref{tab:sys}. Assuming that all these components are independent, they are summed in quadrature to give the totals 0.82\% and 0.87\% for the branching fractions of $D^{+}\to K_{S}^{0}X$ and $D^{0}\to K_{S}^{0}X$,
respectively.

\begin{table}[htbp]
\small
\centering
\caption{Relative systematic uncertainties (in \%) in the branching fraction measurements. A ``...'' denotes that there is no systematic uncertainty. }
\begin{center}
\begin{tabular}{ccccccc}
\hline
\hline
Source                      &$D^{+}\to K_{S}^{0}X$    &$D^{0}\to K_{S}^{0}X$  \\
\hline
$N_{\rm tag}$                  &0.50 &0.50\\
$K_{S}^{0}$ reconstruction     &0.53 &0.53\\
$K_{S}^{0}$ multiplicity       &0.20 &0.28\\
$K_{S}^{0}$ sideband           &0.15 &0.10\\
Quantum-correlation effect     &...  &0.24\\
MC statistics                  &0.25 &0.28\\
Quoted branching fraction      &0.07 &0.07\\
Scale factor $f_{\rm co}$      &0.02 &0.05\\
Total                          &0.82 &0.87\\ \hline
\hline
\end{tabular}
\end{center}
\label{tab:sys}
\end{table}

\begin{table*}[htbp]
\centering
\caption{Comparisons of the  branching fractions of $D^{+(0)}\to K_{S}^{0}X$ measured in this study and Mark-III, BES,  and PDG. For the values from BES, Mark-III, and PDG, we take half of the branching fractions of $D\to K^0/\bar K^0 X$.
${\mathcal B}^{\rm sum}_{\rm exclusive}$ denotes the total branching fractions summing over all known exclusive $D^{+(0)}$ decays involving a $K^0_S$.}
\begin{center}
\begin{tabular}{cccccc}
\hline
\hline
Decay mode    & Mark-III (\%)~\cite{Mark3-KSX}  & BES (\%)~\cite{BESII-KSX}  & PDG (\%)~\cite{pdg2020}  &This study (\%) & ${\mathcal B}^{\rm sum}_{\rm exclusive}$ (\%)\\
\hline
$D^+ \to K_S^{0}X$&$30.60\pm3.25\pm2.15$ &$30.25\pm2.75\pm1.65$&$30.5\pm2.5$  &$32.78\pm0.13\pm0.27$ &$31.68\pm0.32$ \\
$D^0 \to K_S^{0}X$&$22.75\pm2.50\pm1.60$ &$23.80\pm2.40\pm1.50$&$23.5\pm2.0$  &$20.54\pm0.12\pm0.18$ &$18.16\pm0.72$\\

\hline
\hline
\end{tabular}
\end{center}
\label{tagyields}
\end{table*}

\section{Summary}

By analyzing 2.93 fb$^{-1}$ of $e^+e^-$ collision data taken at the center-of-mass energy of  3.773 GeV with the BESIII detector,
the branching fractions of $D^+\to K^0_S X$
and $D^0\to K^0_S X$ are measured
to be $(32.78\pm 0.13\pm 0.27)\%$ and $(20.54\pm 0.12\pm 0.18)\%$, respectively, where the first uncertainties are statistical and the second are systematic.
These results are consistent with  previous measurements by Mark-III~\cite{Mark3-KSX} and BES~\cite{BESII-KSX} within uncertainties, as summarized in Table~\ref{tagyields}.
Compared with the average values from the PDG, the precision of the branching fractions of  $D^+\to K^0_S X$
and $D^0\to K^0_S X$ is improved by factors of 9.0 and 8.1, respectively.
Summing over the branching fractions of the known $D^{+(0)}$ decay modes containing $K^0_S$,
we obtain ${\mathcal B}^{\rm sum}_{\rm exclusive}(D^+\to K^0_SX)=(31.68\pm0.32)\%$ and ${\mathcal B}^{\rm sum}_{\rm exclusive}(D^0\to K^0_SX)=(18.16\pm0.72)\%$.
The differences between the inclusive and exclusive decay branching fractions
are $(1.10\pm0.41)\%$ and $(2.38\pm0.75)\%$ for $D^+$ and $D^0$ decays, respectively.
These results indicate that there may be some missing decay modes involving $K_S^0$ for both $D^+$ and $D^0$ yet to be observed.

\section{Acknowledgement}

The BESIII collaboration thanks the staff of BEPCII and the IHEP computing center for their strong support. This work is supported in part by National Key R\&D Program of China under Contracts Nos. 2020YFA0406400, 2020YFA0406300; National Natural Science Foundation of China (NSFC) under Contracts Nos. 11635010, 11735014, 11835012, 11935015, 11935016, 11935018, 11961141012, 12022510, 12025502, 12035009, 12035013, 12192260, 12192261, 12192262, 12192263, 12192264, 12192265; the Chinese Academy of Sciences (CAS) Large-Scale Scientific Facility Program; Joint Large-Scale Scientific Facility Funds of the NSFC and CAS under Contract No. U1832207, U1932102; the CAS Center for Excellence in Particle Physics (CCEPP); 100 Talents Program of CAS; The Institute of Nuclear and Particle Physics (INPAC) and Shanghai Key Laboratory for Particle Physics and Cosmology; ERC under Contract No. 758462; European Union's Horizon 2020 research and innovation programme under Marie Sklodowska-Curie grant agreement under Contract No. 894790; German Research Foundation DFG under Contracts Nos. 443159800, Collaborative Research Center CRC 1044, GRK 2149; Istituto Nazionale di Fisica Nucleare, Italy; Ministry of Development of Turkey under Contract No. DPT2006K-120470; National Science and Technology fund; National Science Research and Innovation Fund (NSRF) via the Program Management Unit for Human Resources \& Institutional Development, Research and Innovation under Contract No. B16F640076; STFC (United Kingdom); Suranaree University of Technology (SUT), Thailand Science Research and Innovation (TSRI), and National Science Research and Innovation Fund (NSRF) under Contract No. 160355; The Royal Society, UK under Contracts Nos. DH140054, DH160214; The Swedish Research Council; U. S. Department of Energy under Contract No. DE-FG02-05ER41374.

\appendix{
\section{Branching fractions of the known exclusive $D^{+(0)}$ decays involving $K_S^0$}
\label{D0Dpmodes}

Tables~\ref{tab:Dpmodes} and~\ref{tab:D0modes} summarize the branching fractions of the known exclusive $D^+$ and $D^0$ decays involving $K_S^0$, respectively.

\begin{table*}[hbtp]
\centering
%\scriptsize
\footnotesize
\caption{Initial and final states contributing to the inclusive decay $D^+ \to K_S^{0}X$ and the corresponding branching fractions.}
\begin{center}
%\begin{spacing}{0.75}
\begin{tabular}{cccc}
\hline
\hline
Initial state    &$\mathcal{B}$ (\%) &Final state  &$\mathcal{B}$ (\%)   \\
\hline
$K_{S}^{0}\pi^{+}$              &$1.49\pm0.03$ \cite{ref18}  &$K_{S}^{0}\pi^{+}$              &$1.49\pm0.03$    \\
$\bar K^{0}e^{+}\nu_{e}$        &$8.73\pm0.10$ \cite{pdg2020}&$K_{S}^{0}e^{+}\nu_{e}$     &$4.37\pm0.05$        \\
$\bar K^{*0}e^{+}\nu_{e}$       &$5.40\pm0.10$ \cite{pdg2020}&$K_{S}^{0}\pi^{0}e^{+}\nu_{e}$  &$0.89\pm0.02$\\
$\bar K^{0}\mu^{+}\nu_{\mu}$    &$8.76\pm0.19$ \cite{pdg2020}  &$K_{S}^{0}\mu^{+}\nu_{\mu}$  &$4.380\pm0.095$     \\
$\bar K^{*0}\mu^{+}\nu_{\mu}$   &$5.27\pm0.15$ \cite{pdg2020}  &$K_{S}^{0}\pi^{0}\mu^{+}\nu_{\mu}$  &$0.88\pm0.02$\\
$K_{S}^{0}K_{S}^{0}K^{+}$       &$0.254\pm0.013$ \cite{pdg2020}&$K_{S}^{0}K_{S}^{0}K^{+}$       &$0.254\pm0.013$  \\
$K_{S}^{0}K_{L}^{0}K^{+}$       &$0.508\pm0.026$ \cite{pdg2020}&$K_{S}^{0}K_{L}^{0}K^{+}$       &$0.508\pm0.026$  \\

$K_{S}^{0}K_{S}^{0}\pi^{+}$     &$0.270\pm0.013$ \cite{pdg2020}&$K_{S}^{0}K_{S}^{0}\pi^{+}$     &$0.270\pm0.013$  \\
$\bar K^{0}\pi^{+}\pi^{+}\pi^{-}$ &$6.20\pm0.18$ \cite{pdg2020} &$K_{S}^{0}\pi^{+}\pi^{+}\pi^{-}$&$3.10\pm0.09$     \\
$\bar K^{0}K^{+}K^{-}\pi^{+}$   &$0.048\pm0.010$ \cite{pdg2020}&$K_{S}^{0}K^{+}K^{-}\pi^{+}$    &$0.024\pm0.005$  \\
$K_{S}^{0}\pi^{+}\pi^{0}$       &$7.36\pm0.21$ \cite{pdg2020}  &$K_{S}^{0}\pi^{+}\pi^{0}$       &$7.36\pm0.21$   \\
$\bar K^{0}{K^{*}}^{+}$         &$1.738\pm0.182$ \cite{bes3-kskpi0}&$K_{S}^{0}{K^{*}}^{+}$          &$0.869\pm0.091$  \\
$\bar K_{1}^{0}(1270)e^{+}\nu_e$         &$0.231\pm0.030$ \cite{bes3-Dp-K1ev}  &$K_{S}^{0}\pi^{0}\pi^{0}e^{+}\nu_{e}$   &$0.0055\pm0.0007$      \\
$\bar K_{1}^{0}(1270)e^{+}\nu_e$         &$0.231\pm0.030$ \cite{bes3-Dp-K1ev}  &$K_{S}^{0}\pi^{+}\pi^{-}e^{+}\nu_{e}$    &$0.039\pm0.005$      \\
$\bar K_{1}^{0}(1270)\mu^{+}\nu_{\mu}$   &$0.231\pm0.030$ \cite{bes3-Dp-K1ev} &$K_{S}^{0}\pi^{+}\pi^{-}\mu^{+}\nu_{\mu}$    &$0.039\pm0.005$  \\
$\bar K_{1}^{0}(1270)\mu^{+}\nu_{\mu}$   &$0.231\pm0.030$ \cite{bes3-Dp-K1ev} &$K_{S}^{0}\pi^{0}\pi^{0}\mu^{+}\nu_{\mu}$    &$0.0055\pm0.0007$  \\
$K_{S}^{0}\pi^{+}\eta$          &$1.309\pm0.048$ \cite{bes3-etaX}&$K_{S}^{0}\pi^{+}\eta$      &$1.309\pm0.048$      \\
$K_{S}^{0}K^{+}\eta$          &$0.0185\pm0.0050$ \cite{bes3-etaX}&$K_{S}^{0}K^{+}\eta$       &$0.0185\pm0.0050$     \\
$K_{S}^{0}\pi^{+}\pi^{0}\eta$     &$0.12\pm0.03$ \cite{bes3-etaX}  &$K_{S}^{0}\pi^{+}\pi^{0}\eta$ &$0.12\pm0.03$    \\
$K_{S}^{0}K^{+}\pi^{+}\pi^{-}$ &$0.189\pm0.013$ \cite{bes3-D-KKpipi}&$K_{S}^{0}K^{+}\pi^{+}\pi^{-}$&$0.189\pm0.013$ \\
$K_{S}^{0}K^{-}\pi^{+}\pi^{+}$    &$0.227\pm0.013$ \cite{bes3-D-KKpipi}&$K_{S}^{0}K^{-}\pi^{+}\pi^{+}$   &$0.227\pm0.013$   \\
$K_{S}^{0}K^{+}\pi^{0}\pi^{0}$    &$0.058\pm0.013$ \cite{bes3-D-KKpipi} &$K_{S}^{0}K^{+}\pi^{0}\pi^{0}$   &$0.058\pm0.013$  \\
$K_{S}^{0}K_{S}^{0}\pi^{+}\pi^{0}$&$0.134\pm0.021$ \cite{bes3-D-KKpipi} &$K_{S}^{0}K_{S}^{0}\pi^{+}\pi^{0}$ &$0.134\pm0.021$ \\
$K_{S}^{0}K_{L}^{0}\pi^{+}\pi^{0}$&$0.268\pm0.042$ \cite{bes3-D-KKpipi}&$K_{S}^{0}K_{L}^{0}\pi^{+}\pi^{0}$&$0.268\pm0.042$ \\
$\bar K^{0}K^{+}$               &$0.604\pm0.024$ \cite{k0k}&$K_{S}^{0}K^{+}$                &$0.302\pm0.012$   \\
$K_{S}^{0}K^{+}\pi^{0}$         &$0.507\pm0.003$ \cite{k0k} &$K_{S}^{0}K^{+}\pi^{0}$         &$0.507\pm0.003$  \\
$\phi\pi^{+}$                   &$0.570\pm0.015$ \cite{phipi} &$K_{S}^{0}K_{L}^{0}\pi^{+}$     &$0.193\pm0.006$   \\
$\bar K^{0}\pi^{+}\eta^\prime$  &$0.380\pm0.042$ \cite{k0pieta}&$K_{S}^{0}\pi^{+}\eta^\prime$  &$0.190\pm0.021$    \\
$K_{S}^{0}\omega\pi^{+}$        &$0.707\pm0.050$ \cite{ksompi}&$K_{S}^{0}\omega\pi^{+}$      &$0.707\pm0.050$      \\
$K_{S}^{0}\pi^{+}\pi^{0}\pi^{0}$ &$2.2516\pm0.1068$ \cite{kspipi0pi0}&$K_{S}^{0}\pi^{+}\pi^{0}\pi^{0}$ &$2.2516\pm0.1068$  \\
$K_{S}^{0}\pi^{+}\pi^{+}\pi^{-}\pi^{0}$ &$0.4001\pm0.0550$ \cite{kspipi0pi0} &$K_{S}^{0}\pi^{+}\pi^{+}\pi^{-}\pi^{0}$ &$0.4001\pm0.0550$      \\
$K_{S}^{0}\pi^{+}\pi^{0}\pi^{0}\pi^{0}$ &$0.126\pm0.050$ \cite{kspipi0pi0} &$K_{S}^{0}\pi^{+}\pi^{0}\pi^{0}\pi^{0}$ &$0.126\pm0.050$ \\
$\phi\pi^{+}\pi^{0}$            &$0.579\pm0.055$ \cite{phipipi0}&$K_{S}^{0}K_{L}^{0}\pi^{+}\pi^{0}$&$0.196\pm0.019$    \\
\hline
Sum          &...   &...   &$31.68\pm0.32$   \\
\hline
\hline
\end{tabular}
\end{center}
\label{tab:Dpmodes}
\end{table*}

\begin{table*}[hbtp]
\centering
%\scriptsize
\footnotesize
\caption{Initial and final states contributing to the inclusive decay $D^0 \to K_S^{0}X$ and the corresponding branching fractions.
}
\begin{center}
\begin{tabular}{cccc}
\hline
\hline
Initial state     &$\mathcal{B}$  (\%)  &Final state  &$\mathcal{B}$  (\%)    \\
\hline
$K_{S}^{0}\omega$                &$1.11\pm0.12$ \cite{pdg2020}&$K_{S}^{0}\omega$    &$1.11\pm0.12$    \\
$K_{S}^{0}\eta^\prime\pi^{0}$ &$0.252\pm0.027$ \cite{pdg2020}&$K_{S}^{0}\eta^\prime\pi^{0}$&$0.252\pm0.027$    \\
$K^{-}a_{0}^{+}$      &$0.059\pm0.018$ \cite{pdg2020}&$K_{S}^{0}K^{+}K^{-}$   &$0.003\pm0.001$    \\
$K_{S}^{0}\phi$       &$0.444\pm0.015$ \cite{pdg2020}&$K_{S}^{0}\phi$         &$0.444\pm0.015$    \\
$\phi K_{L}^{0}$    &$0.4440\pm0.0304$ \cite{pdg2020}&$K_{S}^{0}K_{L}^{0}K_{L}^{0}$     &$0.151\pm0.010$\\
$\bar K^{*0}\pi^{0}$  &$1.82\pm0.07$ \cite{pdg2020}  &$K_{S}^{0}\pi^{0}\pi^{0}$         &$0.303\pm0.012$      \\
$\phi\gamma$          &$0.00282\pm0.00019$ \cite{pdg2020}&$K_{S}^{0}K_{L}^{0}\gamma$      &$0.0010\pm0.0001$\\
$K_{S}^{0}\pi^{+}\pi^{-}$  &$2.80\pm0.18$ \cite{pdg2020}  &$K_{S}^{0}\pi^{+}\pi^{-}$     &$2.80\pm0.18$      \\
$K_{S}^{0}K^{+}\pi^{-}$    &$0.217\pm0.034$ \cite{pdg2020} &$K_{S}^{0}K^{+}\pi^{-}$      &$0.217\pm0.034$    \\
$K_{S}^{0}K^{-}\pi^{+}$          &$0.33\pm0.05$ \cite{pdg2020}    &$K_{S}^{0}K^{-}\pi^{+}$   &$0.33\pm0.05$          \\
$K_{S}^{0}K_{S}^{0}K^{-}\pi^{+}$  &$0.059\pm0.013$ \cite{pdg2020}&$K_{S}^{0}K_{S}^{0}K^{-}\pi^{+}$   &$0.059\pm0.013$    \\
$K_{S}^{0}K_{L}^{0}K^{-}\pi^{+}$ &$0.1180\pm0.0013$ \cite{pdg2020} &$K_{S}^{0}K_{L}^{0}K^{-}\pi^{+}$ &$0.1180\pm0.0013$\\
$K_{S}^{0}K_{S}^{0}K^{+}\pi^{-}$  &$0.059\pm0.013$ \cite{pdg2020}&$K_{S}^{0}K_{S}^{0}K^{+}\pi^{-}$   &$0.059\pm0.013$    \\
$K_{S}^{0}K_{L}^{0}K^{+}\pi^{-}$ &$0.1180\pm0.0013$ \cite{pdg2020}&$K_{S}^{0}K_{L}^{0}K^{+}\pi^{-}$ &$0.1180\pm0.0013$\\
$K_{S}^{0}\rho^{0}\pi^{+}\pi^{-}$ &$0.11\pm0.07$ \cite{pdg2020}  &$K_{S}^{0}\pi^{+}\pi^{-}\pi^{+}\pi^{-}$ &$0.11\pm0.07$\\
$K^{*-}\pi^{+}\pi^{+}\pi^{-}$     &$0.10\pm0.07$ \cite{pdg2020}   &$K_{S}^{0}\pi^{+}\pi^{-}\pi^{+}\pi^{-}$ &$0.033\pm0.023$\\
$K_{S}^{0}\pi^{+}\pi^{-}\pi^{0}$   &$4.09\pm0.60$ \cite{pdg2020} &$K_{S}^{0}\pi^{+}\pi^{-}\pi^{0}$   &$4.09\pm0.60$\\
$\phi\pi^{+}\pi^{-}$        &$0.02\pm0.01$ \cite{pdg2020}  &$K_{S}^{0}K_{L}^{0}\pi^{+}\pi^{-}$ &$0.0068\pm0.0034$\\
$\bar K^{*0}\gamma$   &$0.042\pm0.007$ \cite{pdg2020}  &$K_{S}^{0}\pi^{0}\gamma$          &$0.007\pm0.001$     \\
$K_{S}^{0}\pi^{0}\eta$            &$1.006\pm0.046$ \cite{bes3-etaX} &$K_{S}^{0}\pi^{0}\eta$             &$1.006\pm0.046$   \\
$K_{S}^{0}K_{S}^{0}\eta$          &$0.0130\pm0.0062$ \cite{bes3-etaX}&$K_{S}^{0}K_{S}^{0}\eta$         &$0.0130\pm0.0062$ \\
$K_{S}^{0}K_{L}^{0}\eta$   &$0.0270\pm0.0124$ \cite{bes3-etaX}  &$K_{S}^{0}K_{L}^{0}\eta$   &$0.0270\pm0.0124$\\
$K_{S}^{0}\pi^{+}\pi^{-}\eta$     &$0.280\pm0.021$ \cite{bes3-etaX} &$K_{S}^{0}\pi^{+}\pi^{-}\eta$      &$0.280\pm0.021$    \\
$K_{S}^{0}\pi^{0}\pi^{0}\eta$     &$0.176\pm0.026$ \cite{bes3-etaX}&$K_{S}^{0}\pi^{0}\pi^{0}\eta$      &$0.176\pm0.026$    \\
$K_{S}^{0}K^{-}\pi^{+}\pi^{0}$    &$0.132\pm0.016$ \cite{bes3-D-KKpipi}&$K_{S}^{0}K^{-}\pi^{+}\pi^{0}$     &$0.132\pm0.016$   \\
$K_{S}^{0}K_{S}^{0}\pi^{+}\pi^{-}$&$0.053\pm0.009$ \cite{bes3-D-KKpipi}&$K_{S}^{0}K_{S}^{0}\pi^{+}\pi^{-}$ &$0.053\pm0.009$   \\
$K_{S}^{0}K_{L}^{0}\pi^{+}\pi^{-}$ &$0.106\pm0.018$ \cite{bes3-D-KKpipi}&$K_{S}^{0}K_{L}^{0}\pi^{+}\pi^{-}$ &$0.106\pm0.018$\\
$K_{S}^{0}K^{-}\pi^{+}\pi^{0}$    &$0.132\pm0.016$ \cite{bes3-D-KKpipi}&$K_{S}^{0}K^{-}\pi^{+}\pi^{0}$     &$0.132\pm0.016$   \\
$K_{S}^{0}K^{+}\pi^{-}\pi^{0}$    &$0.065\pm0.008$ \cite{bes3-D-KKpipi} &$K_{S}^{0}K^{+}\pi^{-}\pi^{0}$     &$0.065\pm0.008$    \\
$\phi\pi^{0}$          &$0.117\pm0.004$ \cite{phipi} &$K_{S}^{0}K_{L}^{0}\pi^{0}$       &$0.040\pm0.001$      \\
$\phi\eta$                       &$0.018\pm0.005$ \cite{phipi} &$K_{S}^{0}K_{L}^{0}\eta$          &$0.006\pm0.001$    \\
$K_{S}^{0}\pi^{0}\omega$          &$0.848\pm0.055$ \cite{ksompi}&$K_{S}^{0}\pi^{0}\omega$           &$0.848\pm0.055$   \\
$K_{S}^{0}\pi^{0}\pi^{0}\pi^{0}$  &$0.591\pm0.042$ \cite{kspipi0pi0}&$K_{S}^{0}\pi^{0}\pi^{0}\pi^{0}$   &$0.591\pm0.042$   \\
$K_{S}^{0}\pi^{+}\pi^{-}\pi^{0}\pi^{0}$ &$0.327\pm0.045$ \cite{kspipi0pi0} &$K_{S}^{0}\pi^{+}\pi^{-}\pi^{0}\pi^{0}$ &$0.327\pm0.045$ \\
$K_{S}^{0}\eta^\prime$              &$0.949\pm0.016$ \cite{ref31}&$K_{S}^{0}\eta^\prime$                &$0.949\pm0.016$    \\
$K_{S}^{0}\pi^{0}$               &$1.240\pm0.022$ \cite{ref31}&$K_{S}^{0}\pi^{0}$                 &$1.240\pm0.022$     \\
${K^{*}}^{-}e^{+}\nu_{e}$        &$2.03\pm0.66$ \cite{k*enu}  &$K_{S}^{0}\pi^{-}e^{+}\nu_{e}$     &$0.68\pm0.22$     \\
$\bar K^{0}\pi^{-}e^{+}\nu_{e}$  &$0.158\pm0.034$ \cite{k*enu} &$K_{S}^{0}\pi^{-}e^{+}\nu_{e}$     &$0.079\pm0.017$    \\
${K^{*}}^{-}\mu^{+}\nu_{\mu}$    &$1.91\pm0.62$ \cite{k*enu}  &$K_{S}^{0}\pi^{-}\mu^{+}\nu_{\mu}$ &$0.63\pm0.20$     \\
${K_{1}}^{-}(1270)e^{+}\nu_e$          &$0.109\pm0.018$ \cite{k1enu} &$K_{S}^{0}\pi^{0}\pi^{-}e^{+}\nu_e$&$0.026\pm0.004$    \\
${K_{1}}^{-}(1270)\mu^{+}\nu_{\mu}$    &$0.109\pm0.018$ \cite{k1enu} &$K_{S}^{0}\pi^{0}\pi^{-}\mu^{+}\nu_{\mu}$ &$0.026\pm0.004$    \\
$K_{S}^{0}K_{S}^{0}$             &$0.0170\pm0.0015$ \cite{ksks}&$K_{S}^{0}K_{S}^{0}$              &$0.0170\pm0.0015$  \\
$K_{S}^{0}K_{S}^{0}K_{S}^{0}$    &$0.072\pm0.007$ \cite{ksks} &$K_{S}^{0}K_{S}^{0}K_{S}^{0}$     &$0.072\pm0.007$    \\
$K_{S}^{0}K_{L}^{0}K_{L}^{0}$   &$0.216\pm0.021$ \cite{ksks} &$K_{S}^{0}K_{L}^{0}K_{L}^{0}$ &$0.216\pm0.021$\\
$K_{S}^{0}K_{S}^{0}K_{L}^{0}$   &$0.216\pm0.021$ \cite{ksks} &$K_{S}^{0}K_{S}^{0}K_{L}^{0}$ &$0.216\pm0.021$\\

\hline
Sum            &...   &...  &$18.16\pm0.72$ \\
\hline
\hline
\end{tabular}
\end{center}
\label{tab:D0modes}
\end{table*}

}


\begin{thebibliography}{**}

\bibitem{Mark3-KSX}
D.~Coffman {\it et al}. (Mark-III Collaboration),
\href{https://doi.org/10.1016/0370-2693(91)91718-B}{Phys. Lett. B {\bf 263}, 135 (1991).}

\bibitem{BESII-KSX}
M.~Ablikim {\it et al}. (BES Collaboration),
\href{https://doi.org/10.1016/j.physletb.2006.10.057}{Phys. Lett. B {\bf 643}, 246 (2006).}

\bibitem{pdg2020}
R.~L.~Workman {\it et al.} (Particle Data Group),
\href{http://pdglivetest.lbl.gov/Viewer.action}{Prog. Theor. Exp. Phys. {\bf 2022}, 083C01 (2022).}


\bibitem{lum_bes3}
M.~Ablikim {\it et al.} (BESIII Collaboration),
\href{https://iopscience.iop.org/article/10.1088/1674-1137/37/12/123001}{Chin. Phys. C {\bf 37}, 123001 (2013);}
\href{https://doi.org/10.1016/j.physletb.2015.11.043}{Phys. Lett. B {\bf 753}, 629 (2016).}


\bibitem{BESIII}
M.~Ablikim {\it et al.} (BESIII Collaboration),
\href{https://doi.org/10.1016/j.nima.2009.12.050}{Nucl. Instrum. Meth. A {\bf 614}, 345 (2010).}

\bibitem{Yu:IPAC2016-TUYA01}
C.~H.~Yu {\it et al.},
\href{doi:10.18429/JACoW-IPAC2016-TUYA01}{Proceedings of IPAC2016, Busan, Korea, 2016.}

\bibitem{Ablikim:2019hff}
M.~Ablikim {\it et al.} (BESIII Collaboration),
\href{https://doi.org/10.1088/1674-1137/44/4/040001}{Chin. Phys. C {\bf 44}, 040001 (2020).}


\bibitem{detvis}
K.~X.~Huang, {\it et al.},
\href{https://doi.org/10.1007/s41365-022-01133-8}{Nucl.\ Sci.\ Tech. {\bf 33}, 142 (2022).}


\bibitem{geant4}
S.~Agostinelli {\it et al.} (GEANT4 Collaboration),
\href{https://doi.org/10.1016/S0168-9002(03)01368-8}{Nucl. Instrum. Meth. A {\bf 506}, 250 (2003).}

\bibitem{kkmc}
S.~Jadach, B.~F.~L.~Ward, and Z.~Was,
\href{https://linkinghub.elsevier.com/retrieve/pii/S0010465500000485}{ Comp. Phys. Commu. {\bf 130}, 260 (2000);} \href{https://journals.aps.org/prd/abstract/10.1103/PhysRevD.63.113009}{Phys. Rev. D {\bf 63}, 113009 (2001).}

\bibitem{evtgen}
D.~J.~Lange,
\href{https://doi.org/10.1016/S0168-9002(01)00089-4} {Nucl. Instrum. Meth. A {\bf 462}, 152 (2001);}
R.~G.~Ping,
\href{https://doi.org/10.1088/1674-1137/32/8/001}{Chin. Phys. C {\bf 32}, 599 (2008).}

\bibitem{lundcharm}
J.~C.~Chen, G.~S.~Huang, X.~R.~Qi, D.~H.~Zhang, and Y.~S.~Zhu,
\href{https://journals.aps.org/prd/abstract/10.1103/PhysRevD.62.034003}{Phys. Rev. D {\bf 62}, 034003 (2000);}
R.~L.~Yang, R.~G.~Ping and H.~Chen,
\href{https://iopscience.iop.org/article/10.1088/0256-307X/31/6/061301}{Chin. Phys. Lett. 31, 061301 (2014).}

\bibitem{photos}
E.~Richter-Was,
\href{https://doi.org/10.1016/0370-2693(93)90062-M`}{Phys. Lett. B {\bf 303}, 163 (1993).}

\bibitem{bes3-kspipipi}
M.~Ablikim {\it et al}. (BESIII Collaboration),
\href{https://journals.aps.org/prd/pdf/10.1103/PhysRevD.100.072008}{Phys.~Rev. D {\bf 100}, 072008 (2019).}

\bibitem{bes3-D-KKpipi}
M.~Ablikim {\it et al.} (BESIII Collaboration),
\href{https://journals.aps.org/prd/abstract/10.1103/PhysRevD.102.052006}{Phys. Rev. D {\bf 102},  052006  (2020).}

\bibitem{ksompi}
M.~Ablikim {\it et al}. (BESIII Collaboration),
\href{https://journals.aps.org/prd/abstract/10.1103/PhysRevD.105.032009}{Phys. Rev. D {\bf 105}, 032009 (2022).}

\bibitem{lanxing}
M.~Ablikim {\it et al.} (BESIII Collaboration),
\href{https://arxiv.org/abs/2205.14031}{arXiv:2205.14031}.

\bibitem{Li:2021iwf}
H.~B.~Li and X.~R.~Lyu,
\href{https://doi.org/10.1093/nsr/nwab181}{Natl. Sci. Rev. {\bf 8}, nwab181 (2021).}

\bibitem{refcp1}
T.~Evans {\it et al.},
\href{https://doi.org/10.1016/j.physletb.2016.04.037}{Phys. Lett. B {\bf 757}, 520 (2016);}
\href{https://doi.org/10.1016/j.physletb.2016.11.021}{{\bf 765}, 402(E) (2017).}

\bibitem{refcp2}
M.~Ablikim {\it et al.} (BESIII Collaboration),
\href{https://journals.aps.org/prd/abstract/10.1103/PhysRevD.100.072006}{Phys. Rev. D {\bf 100}, 072006 (2019).}

\bibitem{A1}
Heavy Flavor Averaging Group (HFLAV),
\href{http://www.slac.stanford.edu/xorg/hflav/charm/}{(http://www.slac.stanford.edu/xorg/hflav/charm/).}

\bibitem{refcp3}
T.~Gershon, J.~Libby, and G.~Wilkinson,
\href{https://doi.org/10.1016/j.physletb.2015.08.063} {Phys. Lett. B {\bf 750}, 338 (2015).}

\bibitem{refcp4}
T.~Evans {\it et al.},
\href{https://doi.org/10.1016/j.physletb.2016.04.037}{Phys. Lett. B {\bf 757}, 520 (2016);}
\href{https://doi.org/10.1016/j.physletb.2016.11.021}{{\bf 765}, 402(E) (2017).}


\bibitem{bes3-pimuv}
M.~Ablikim {\it et al.} (BESIII Collaboration),
\href{https://journals.aps.org/prl/abstract/10.1103/PhysRevLett.121.171803}{Phys. Rev. Lett. {\bf 121}, 171803 (2018).}

\bibitem{bes3-etaetapi}
M.~Ablikim {\it et al.} (BESIII Collaboration),
\href{https://journals.aps.org/prd/pdf/10.1103/PhysRevD.101.052009}{Phys. Rev. D {\bf 101}, 052009 (2020).}

\bibitem{bes3-etaX}
M.~Ablikim {\it et al.} (BESIII Collaboration),
\href{https://dx.doi.org/10.1103/PhysRevLett.124.241803}{Phys. Rev. Lett. {\bf 124}, 241803 (2020).}

\bibitem{deltakpi}
M.~Ablikim {\it et al.} (BESIII Collaboration),
\href{https://doi.org/10.1016/j.physletb.2014.05.071}{Phys. Lett. B {\bf 734}, 227 (2014).}

\bibitem{ARGUS}
H.~Albrecht {\it et al.} (ARGUS Collaboration),
\href{https://doi.org/10.1016/0370-2693(90)91293-K}{Phys. Lett. B {\bf 241}, 278 (1990).}

\bibitem{kso_reconstruction}
M.~Ablikim {\it et al.} (BESIII Collaboration),
\href{https://journals.aps.org/prd/abstract/10.1103/PhysRevD.92.112008}{Phys. Rev. D {\bf 92}, 112008 (2015).}

\bibitem{ref18}
C.~Patrignani {\it et al}. (Particle Data Group),
\href{https://doi.org/10.1088/1674-1137/40/10/100001}{Chin. Phys. C 40, 100001 (2016).}

\bibitem{bes3-kskpi0}
M.~Ablikim {\it et al}. (BESIII Collaboration),
\href{https://journals.aps.org/prd/abstract/10.1103/PhysRevD.104.012006}{Phys. Rev. D {\bf 104}, 012006 (2021).}

\bibitem{bes3-Dp-K1ev}
M.~Ablikim {\it et al.} (BESIII Collaboration),
\href{https://journals.aps.org/prl/abstract/10.1103/PhysRevLett.123.231801}{Phys. Rev. Lett. {\bf 123}, 231801 (2019).}

\bibitem{k0k}
M.~Ablikim {\it et al}. (BESIII Collaboration),
\href{https://journals.aps.org/prd/abstract/10.1103/PhysRevD.99.032002}{Phys. Rev. D {\bf 99}, 032002  (2019).}

\bibitem{phipi}
M.~Ablikim {\it et al}. (BESIII Collaboration),
\href{https://doi.org/10.1016/j.physletb.2019.135017}{Phys. Lett. B {\bf 798}, 135017  (2019).}

\bibitem{k0pieta}
M.~Ablikim {\it et al}. (BESIII Collaboration),
\href{https://journals.aps.org/prd/abstract/10.1103/PhysRevD.98.092009}{Phys. Rev. D {\bf 99}, 092009 (2018).}

\bibitem{kspipi0pi0}
M.~Ablikim {\it et al}. (BESIII Collaboration), \href{https://arxiv.org/abs/2205.14031}{arXiv:2205.14031v1.}

\bibitem{phipipi0}
M.~Ablikim {\it et al}. (BESIII Collaboration),
``Measurement of the branching fraction of $D^{+}\to \phi\pi^+\pi^0$'', publication in preparation.

\bibitem{ref31}
M.~Ablikim {\it et al}. (BESIII Collaboration),
\href{https://journals.aps.org/prd/abstract/10.1103/PhysRevD.97.072004}{Phys. Rev. D {\bf 97}, 072004 (2018).}

\bibitem{k*enu}
M.~Ablikim {\it et al}. (BESIII Collaboration),
\href{https://journals.aps.org/prd/abstract/10.1103/PhysRevD.99.011103}{Phys. Rev. D {\bf 99}, 011103 (2019).}

\bibitem{k1enu}
M.~Ablikim {\it et al}. (BESIII Collaboration),
\href{https://journals.aps.org/prl/abstract/10.1103/PhysRevLett.127.131801}{Phys. Rev. Lett. {\bf 127}, 131801 (2021).}

\bibitem{ksks}
M.~Ablikim {\it et al}. (BESIII Collaboration),
\href{https://doi.org/10.1016/j.physletb.2016.12.020}{Phys. Lett. B {\bf 765}, 231  (2017).}


\end{thebibliography}
\end{document}